\documentclass[aps,prd,reprint,groupedaddress,showpacs,eqsecnum,floatfix]{revtex4-1}

\usepackage{amsmath}
\usepackage{amssymb}

\begin{document}

\title{Physical meaning of the conserved quantities on anti-de Sitter geodesics}

\author{Ion I. Cot\u aescu\\ {\small \it The West 
                 University of Timi\c soara,}\\
   {\small \it V. P\^ arvan Ave. 4, RO-1900 Timi\c soara, Romania}}

\date{\today}
%\draft
%\maketitle

\begin{abstract}
The geodesic motion on anti-de Sitter spacetimes is studied pointing out how the trajectories are determined by the ten independent conserved quantities associated to the specific $SO(2,3)$ isometries of these manifolds.  The new result is that there are two conserved $SO(3)$ vectors which play the same role as the Runge-Lenz vector of the Kepler problem, determining the major and minor semiaxes of the ellipsoidal anti-de Sitter geodesics.   


\end{abstract}

\maketitle
%\newpage

\section{Introduction}
\
In general relativity the anti-de Sitter (AdS) spacetime is the only maximally symmetric manifold which does not have translations \cite{W}. Therefore, the geodesic motion is oscillatory around the origins of the central charts (i. e. static and spherically symmetric) with ellipsoidal closed trajectories. This particular behavior was studied  by many authors \cite{Cal,Haw,Grif,P1,P2,L,C1} from long time and is still of actual interest  \cite{SG,Biz} such that today we are able to understand the principal features of these spacetimes.

However, there are some technical details which were less studied as the transformation of the conserved quantities under the AdS isometries that was only tangentially considered in applications \cite{L,Biz}. Moreover, the physical meaning of the conserved quantities along the AdS geodesics is not elucidated completely since, apart from the energy and angular momentum currently used in integrating the geosesic equations on central  backgrounds, there are other quantities whose meaning remains obscure.

The four-dimensional AdS spacetime has ten independent Killing vectors corresponding to its specific $SO(2,3)$ isometries. These give rise to ten independent conserved quantities along the time-like geodesics among them one identifies the energy and angular momentum components. The other six can be seen as the components of two usual $SO(3)$ vectors (as we explain later) but whose interpretation is not yet established. 

In this short paper we would like to show that these conserved quantities play the same kinematical role as the Runge-Lenz vector \cite{Ru,Le} in the Keplerian motion, determining the major and minor semiaxes of the ellipsoidal time-like geodesics of the AdS spacetime.  

However, despite of this analogy, there is a major difference since the components of the Runge-Lenz vector are  quadratic forms in velocity (or momentum), generated by St\" ackel-Killing tensors \cite{St}, while  our conserved quantities are linear forms given by the Killing vectors associated to isometries. In general, the St\" ackel-Killing tensors are related to the so called {\em hidden} symmetries of some special systems as the Keplerian one or the Kerr \cite{Kerr,Car} or Taub-NUT \cite{G1,G2,G3} geometries.  Other similar geometries were investigated looking for such conserved quantities \cite{2D}.      

We start in the second section presenting the conserved quantities generated by the AdS  Killing vectors in the standard central chart with Cartesian coordinates. A convenient solution of the geosesic equations is given in the next section where we consider another central chart suitable for physical interpretations \cite{P1,P2}. In this manner, we arrive to our new result writing down the components of the conserved vectors determining the semiaxes of the time-like geodesics.  In the last section we present our concluding remarks.    
      
\section{AdS conserved quantities}  

The AdS spacetime $(M,g)$ is defined as the hyperboloid of radius $1/\omega$ embedded  in the five-dimensional pseudo-Euclidean spacetime $(M^5,\eta^5)$ of metric $\eta^5={\rm diag}(1,1,-1,-1,-1)$ where we consider the Cartesian coordinates $z^A$  ($A,\,B,...= -1,0,1,2,3$). The local charts on $M$, of arbitrary coordinates $x^{\mu}$ ($\alpha,...\mu,\nu...=0,1,2,3$),  have to be introduced giving the functions $z^A(x)$ which solve the hyperboloid equation,
\begin{equation}\label{hip}
\eta^5_{AB}z^A(x) z^B(x)=\frac{1}{\omega^2}\,.
\end{equation}
Thus we may consider the local central chart $\{t,\vec{x}\}$ on $M$, with Cartesian spaces coordinates $x^i$ ($i,j,k,...=1,2,3$), defined by
\begin{eqnarray}
z^{-1}(x)&=&\frac{1}{\omega}\chi(r)\cos(\omega t)\,,\\
z^0(x)&=&\frac{1}{\omega}\chi(r)\sin(\omega t)\,,\\
z^i(x)&=&x^i \,, \label{Zx}
\end{eqnarray}
where we denote $r=|\vec{x}|$ and $\chi(r)=\sqrt{1+\omega^2 r^2}$.  Hereby we obtain the line element,  
\begin{eqnarray}
ds^{2}&=&\eta^5_{AB}dz^A(x_c)dz^B(x_c)\nonumber\\
&=&\chi(r)^2 dt^{2}-\left[\delta_{ij}-\frac{\omega^2 x^{i}x^{j}}{\chi(r)^2}\right]dx^{i}dx^{j}\,,\label{line1}
\end{eqnarray}
laying out an obvious central symmetry (under space rotations and time translation). In the chart $\{t,r,\theta,\phi\}$ with spherical coordinates, canonically associated to the Cartesian ones $x^i$, the line element reads
\begin{equation}\label{line2}
ds^2=\chi(r)^2 dt^2-\frac{dr^2}{\chi(r)^2}-r^2(d\theta^2+\sin ^2 \theta\, d\phi^2)\,.
\end{equation}

However, the symmetries of these manifolds are more complicated since their isometries are given by the group $SO(2,3)$ which leave invariant the metric $\eta^5$ of the embedding manifold $(M^5,\eta^5)$ and implicitly Eq. (\ref{hip}). Therefore, given a local chart $\{x\}$  defined by the functions $z=z(x)$, each transformation ${\frak g}\in SO(2,3)$ defines the isometry $x\to x'=\phi_{\frak g}(x)$ derived from the system of equations $z[\phi_{\frak g}(x)]={\frak g}z(x)$. For these transformations we adopt the parametrization
\begin{equation}
{\frak g}(\xi)=\exp\left(-\frac{i}{2}\,\xi^{AB}{\frak S}_{AB}\right)\in SO(2,3) 
\end{equation}
with skew-symmetric parameters, $\xi^{AB}=-\xi^{BA}$,  and the covariant generators ${\frak S}_{AB}$ of the fundamental representation of the $so(2,3)$ algebra carried by $M^5$ that have the matrix elements, 
\begin{equation}
({\frak S}_{AB})^{C\,\cdot}_{\cdot\,D}=i\left(\delta^C_A\, \eta_{BD}^5
-\delta^C_B\, \eta_{AD}^5\right)\,.
\end{equation}
In general, these isometries are not linear transformations apart the rotations ${\frak r}\in SO(3)\subset SO(2,3)$ that may transform linearly the Cartesian coordinates as  $x^i\to \phi^i_{\frak r}(x)=R_{ij}x^j$, but only when these are proportional with $z^i$ as in Eq. (\ref{Zx}). 

The Killing vectors associated to these isometries  are defined (up to a multiplicative constant)  as 
\begin{equation}
k_{(AB)\,\mu}=z_A\partial_{\mu} z_B -z_B\partial_{\mu} z_A,,\quad  z_A=\eta^5_{AC}z^C\,,
\end{equation}
recovering the components given in Ref. \cite{ES}. With their help one can define the conserved quantities of the general form  ${\cal K}_{(AB)}=mk_{(AB)\,\mu} u^{\mu}$, depending on the four-velocity $u^{\mu}=\frac{dx^{\mu}}{ds}$. These quantities transform under isometries $x\to x'=\phi_{\frak g}(x)$ as the components of a five-dimensional skew-symmetric tensor on $(M^5,\eta^5)$ according to the rule 
\begin{equation}\label{KAB}
{\cal K}_{(AB)}'={\frak g}_{A\,\cdot}^{\cdot\,C}\,{\frak g}_{B\,\cdot}^{\cdot\,D}\,{\cal K}_{(CD)}\,,
\end{equation}
that can be put in the matrix form ${\cal K}'=\overline{\frak g}\,{\cal K}\,\overline{\frak g}^T$ where we must use the adjoint transformation matrix $\overline{\frak g}=\eta^5\,{\frak g}\,\eta^5$. Thus we can verify that all the conserved quantities carrying space indices ($i,j,...$) transform alike under rotations as $SO(3)$ vectors or tensors. Moreover, the condition  $z^i\propto x^i$ fixes the same (common)  three-dimensional basis in  both the  Cartesian charts, of  $M^5$ and respectively $M$.  Then we say that the $SO(3)$ symmetry is {\em global} \cite{ES}  and we may use the vector notation for the conserved quantities as well as for the local Cartesian  coordinates of $M$. 

With these preparations we may  introduce ten independent conserved quantities for any massive mobile of mass $m$ freely falling on AdS background. In the central chart with Cartesian coordinates we define the energy
\begin{equation}
E=m\omega k_{(-1,0)\,\mu}u^{\mu}=m\chi(r)^2 u^0\,,
\end{equation}
and the angular momentum  components
\begin{equation}
L_i=m\frac{1}{2}\varepsilon_{ijk}k_{(j,k)\,\mu}u^{\mu}=m\varepsilon_{ijk}x^j u^k\,,
\end{equation}
that have the traditional  physical meaning. In addition, we have two more conserved vectors, $\vec{K}$ and $\vec{N}$, having the components
\begin{eqnarray}
K_i&=&mk_{(i,0)\,\mu}u^{\mu}=m\left[-x^i\chi(r) u^0 \cos(\omega t)\right.\nonumber\\
&&+\left.\left(\frac{\chi(r)}{\omega} u^i-\frac{\omega}{\chi(r)} x^i x^j u^j\right)\sin(\omega t)\right] \,,\\
N_i&=&mk_{(i,-1)\,\mu}u^{\mu}=m\left[x^i\chi(r) u^0 \sin(\omega t)\right.\nonumber\\
&&\left.+\left(\frac{\chi(r)}{\omega} u^i-\frac{\omega}{\chi(r)} x^i x^j u^j\right)\cos(\omega t)\right] \,,
\end{eqnarray} 
and the following obvious properties
\begin{equation}
\vec{K}\cdot \vec{L}=\vec{N}\cdot \vec{L}=0\,, \quad \vec{K}\land \vec{N}=-\frac{E}{\omega} \vec{L}\,.
\end{equation}
Moreover, the general invariant corresponding to the first Casimir operator of the $so(2,3)$ algebra reads now
\begin{equation}
E^2+\omega^2\left({\vec{L}\,}^2 -{\vec{K}\,}^2-{\vec{N}\,}^2\right)=m^2 u^2\,.
\end{equation}

Other new properties of $\vec{K}$ and $\vec{N}$, representing the principal results reported here, may be derived only by using explicitly the geodesic equation.

\section{Time-like geodesics}

In what follows, we restrict ourselves to the time-like case when $u^2=1$  considering the new central chart $\{\tilde x\}=\{t,\rho,\theta,\phi\}$ resulted after the substitution \cite{P1,P2}
\begin{equation}
r=\frac{\rho}{\tilde\chi(\rho)}\,, \quad \tilde\chi(\rho)=\sqrt{1-\omega^2\rho^2}\,,
\end{equation}
where $0\le \rho<\frac{1}{\omega}$. Then the embedding equations become
\begin{eqnarray}
z^{-1}(\tilde x)&=&\frac{1}{\omega\tilde\chi(\rho)}\cos(\omega t)\,,\\
z^0(\tilde x)&=&\frac{1}{\omega\tilde\chi(\rho)}\sin(\omega t)\,,\nonumber\\
z^1(\tilde x)&=&\frac{\rho}{\tilde\chi(\rho)}\sin \theta \cos \phi \,,\\
z^2(\tilde x)&=&\frac{\rho}{\tilde\chi(\rho)}\sin \theta \sin \phi \,,\\
z^3(\tilde x)&=&\frac{\rho}{\tilde\chi(\rho)}\cos \theta
\end{eqnarray}
while the line element reads
\begin{equation}\label{line3}
ds^2=\frac{1}{\tilde\chi(\rho)^2}\left[ dt^2-\frac{d\rho^2}{\tilde\chi(\rho)^2}-\rho^2(d\theta^2+\sin ^2 \theta\, d\phi^2)\right]\,.
\end{equation}  
In this chart  the components of the four-velocity are denoted as $\tilde u^{\mu}=\frac{d\tilde x^{\mu}}{ds}$.

For integrating the geodesic equations and deriving the conserved quantities  it is convenient to take the angular momentum along the third axis,  $\vec{L}=(0,0,L)$, restricting the motion in the equatorial plane, with $\theta=\frac{\pi}{2}$  and $\tilde u^{\theta}=0$. Then the conserved quantities take the form
\begin{eqnarray}
E&=&\frac{m }{\tilde{\chi}^2}\tilde u^t\,,\label{conE}\\
L&=&\frac{m \rho^2 }{\tilde{\chi}^2}\tilde u^{\phi}\,,\label{conL}\\
K_1&=&\frac{m }{\omega\tilde{\chi}^2}\left(-\omega \rho \tilde u^t\cos\omega t\cos\phi \right. \nonumber\\
&&\left.+\tilde u^{\rho}\sin\omega t\cos\phi -\rho \tilde u^{\phi}\sin\omega t\sin\phi\right)\,,\label{K1}\\
K_2&=&\frac{m }{\omega\tilde{\chi}^2}\left(-\omega \rho \tilde u^t\cos\omega t\sin\phi \right. \nonumber\\
&&\left.+\tilde u^{\rho}\sin\omega t\sin\phi +\rho \tilde u^{\phi}\sin\omega t\cos\phi\right)\,,\\
N_1&=&\frac{m }{\omega\tilde{\chi}^2}\left(\omega \rho \tilde u^t\sin\omega t\cos\phi \right. \nonumber\\
&&\left.+\tilde u^{\rho}\cos\omega t\cos\phi -\rho \tilde u^{\phi}\cos\omega t\sin\phi\right)\,,\\
N_2&=&\frac{m }{\omega\tilde{\chi}^2}\left(\omega \rho \tilde u^t\sin\omega t\sin\phi \right. \nonumber\\
&&\left.+\tilde u^{\rho}\cos\omega t\sin\phi +\rho \tilde u^{\phi}\cos\omega t\cos\phi\right)\,,\label{N2}
\end{eqnarray}
and $K_3=N_3=0$. 

Furthermore, from Eqs. (\ref{conE}) and (\ref{conL}) and the identity $\tilde u^2=1$ we obtain  the radial component 
\begin{equation}\label{ur}
\tilde u^{\rho}=\tilde\chi(\rho)^2\left[\frac{E^2}{m^2} \tilde\chi(\rho)^2+\frac{\omega^2 L^2}{m^2}-\frac{L^2}{m^2\rho^2}-1\right]^{\frac{1}{2}}\,,
\end{equation}
deriving the following prime integrals
\begin{eqnarray}
\left(\frac{d\rho}{dt}\right)^2+\omega^2\rho^2+\frac{L^2}{E^2\rho^2}&=&1+\frac{\omega^2 L^2}{E^2}-\frac{m^2}{E^2}\,,\\
\frac{d\phi}{dt}&=&\frac{L}{E\rho^2}\,,
\end{eqnarray}
that have to be integrated.  The solution is known \cite{L,Biz} and can be written with our actual notations  as \cite{C1} 
\begin{eqnarray}
\rho(t)&=&\left[\kappa_1+\kappa_2\cos 2\omega (t-t_0)\right]^{\frac{1}{2}}\,,\label{GEO1}\\
\phi(t)&=&\phi_0+{\rm arctan} \left[\sqrt{\frac{\kappa_1-\kappa_2}{\kappa_1+\kappa_2}}\tan\omega(t- t_0)\right],\label{GEO2}
\end{eqnarray}
where
\begin{eqnarray}
\kappa_1&=&\frac{\omega^2 L^2+E^2-m^2}{2\omega^2 E^2}\,,\\
\kappa_2&=&\frac{1}{2\omega^2 E^2}\left[(E+m)^2-\omega^2 L^2\right]^{\frac{1}{2}} \left[(E-m)^2-\omega^2 L^2\right]^{\frac{1}{2}}\,,\nonumber\\
~
\end{eqnarray}
such that
\begin{equation}
\kappa_1^2-\kappa_2^2=\frac{L^2}{\omega^2 E^2}\,.
\end{equation}
The integration constants $t_0$ and $\phi_0$ determine the initial position of the mobile and implicitly of its trajectory.

Hereby we recover the well-known behavior of the time-like geodesic motion which is oscillatory with frequency $\omega$ having a closed trajectory of an ellipsoidal form in the domain $\rho\in [\rho_{\rm min}, \rho_{\rm max}]$ where
\begin{equation}
\rho_{\rm min}=\sqrt{\kappa_1-\kappa_2}\,, \quad \rho_{\rm max}=\sqrt{\kappa_1+\kappa_2}\,,
\end{equation}
satisfy $\rho_{\rm min}\le\rho_{\rm max}<\frac{1}{\omega}$ for any values of $E$ and $L$. 

For calculating the components of $\vec{K}$ and $\vec{N}$  we set  $t=t_0=0$ fixing the moment when the mobile reaches the aphelion $(\rho_{\rm max}, \phi_0)$ where $\tilde u^{\rho}=0$ as it results from Eq. (\ref{ur}). Then,  by using  Eqs. (\ref{conE})-(\ref{N2}) we obtain:
\begin{eqnarray}
\vec{K}&=&E\rho_{\rm max}\left(\cos \phi_0, \sin \phi_0, 0\right)\,,\\
\vec{N}&=&E\rho_{\rm min}\left(\sin \phi_0, -\cos \phi_0, 0\right)\,.
\end{eqnarray}
The conclusion is that these vectors play a similar role as a Runge-Lenz one  in the sense that they are orthogonal to each other being oriented along the semiaxes of the ellipsoidal trajectory. More specific, the vector $\frac{\vec{K}}{E}$ belongs to the major semiaxis while $\frac{\vec{N}}{E}$ lays over the minor one. 

However, the analogy stops here since, as mentioned before, the components of our conserved vectors are linear forms  in four-velocity while the genuine Runge-Lenz components of the Keplerian motion or of the Taub-NUT geodesics are quadratic forms given by  St\" ackel-Killing tensors.  

\section{Concluding remarks}

We succeeded here to find the physical meaning of all the ten independent conserved quantities of the time-like geodesics on AdS manifolds, including $\vec{K}$ and $\vec{N}$.  It is remarkable that  all these quantities depend only on $E$ and $\vec{L}$ and the initial condition $\phi(t_0)=\phi_0$. This situation is similar to that of the de Sitter spacetimes where the conserved quantities are determined only by the conserved momentum and the initial condition \cite{GRG,C2}. 

The difference is that on de Sitter backgrounds we have translations which are forbidden in the AdS geometry  where all the geodesics are oscillatory motions around the origin of a central chart. For this reason the effect of the AdS isometries may be less relevant even though their study is very difficult in central charts where only the $SO(3)$ symmetry is global. Nevertheless, the example  discussed in  Ref. \cite{Biz} indicates that  the problem of  isometries  transforming  geodesics and their conserved quantities remains open for further investigations with the hope of finding new effective mathematical  methods with physical impact.

\end{document}